\documentclass[sigconf,english,nonacm]{acmart}

\acmConference[MSR 2024]{21st International Conference on Mining Software Repositories}{April 2024}{Lisbon, Portugal}

\AtBeginDocument{}

\setcopyright{acmcopyright}
\copyrightyear{2024}
\acmYear{2024}
\acmDOI{XXXXXXX.XXXXXXX}

\acmPrice{??.00}
\acmISBN{??}

\usepackage{tikz,siunitx}
\usetikzlibrary{backgrounds,arrows.meta}
\tikzset{
   start/.style={fill,minimum size=1ex,circle},
   block/.style={rounded corners=4pt,fill=white,thick,minimum width=3em,minimum height=2em,align=center,font=\sffamily\small,inner sep=.55em,draw=black},
   flow/.style={-Kite,line cap=round,rounded corners=1.5pt},
   artifact/.style={draw,rounded corners=0.5pt,text width=2cm,minimum height=2.15\baselineskip,fill=#1!40!white,align=center},
   artifact/.default=lightgray,
   T/.style={font=\sffamily\scriptsize,darkgray}
}

\def\CRANFiles{358989}
\def\SocialFiles{4230}

\def\ProcessedCRANFiles{357624}
\def\ProcessedCRANTestFiles{78416}
\def\ProcessedCRANExampleFiles{4638}
\edef\ProcessedCRANDefaultFiles{\the\numexpr\ProcessedCRANFiles-\ProcessedCRANTestFiles-\ProcessedCRANExampleFiles\relax}
\def\ProcessedSocialFiles{4083}

\def\AllCallsCRAN{25534260}
\def\AllCallsCRANTest{5615713}

\def\AllCallsSocial{943788}
\def\cpp{C\kern-1.5pt\smash{\raisebox{.33pt}{++}}\xspace}
\usepackage{xfp}

\def\PrintResultOfThe#1#2#3#4{%
   \edef\result{\fpeval{round(#1,#2)}}%
   \edef\chk{\fpeval{\result=0?1:0}}%
   \ifnum\chk=1\relax
      \edef\numcmpres{\fpeval{1/(10^#2)}}%
      \qty{< \numcmpres}{#4}%
   \else
      #3\relax
   \fi
}

\newcommand*\TheFrac[3][2]{%
   \edef\result{\fpeval{(#2)/(#3)}}%
   \PrintResultOfThe{\result}{#1}{\num{\result}}{}%
}
\newcommand*\TheFracIncrease[3][2]{%
   \edef\result{\fpeval{(#2)/(#3)-1}}%
   \PrintResultOfThe{\result}{#1}{\num{\result}}{}%
}
\newcommand*\ThePercent[3][2]{%
   \edef\fst{\fpeval{#2}}%
   \edef\snd{\fpeval{#3}}%
   \edef\result{\fpeval{\fst/\snd * 100}}%
   \PrintResultOfThe{\result}{#1}{\ifdim\result sp=100sp\relax\ifdim\fst sp=\snd sp\else\(\approx\)\fi\fi\qty{\fpeval{\result}}{\percent}}{\percent}%
}

\newcommand*\ThePercentIncrease[3][2]{%
    \edef\fst{\fpeval{#2}}%
    \edef\snd{\fpeval{#3}}%
   \edef\result{\fpeval{round(\fst/\snd * 100 - 100,#1)}}%
   \edef\chk{\fpeval{\result=0?1:0}}%
   \ifnum\chk=1\relax\ifdim\fst sp=0sp\else\(\approx\)\fi\else
   \ifdim\result sp=100sp\relax\ifdim\fst sp=\snd sp\else\(\approx\)\fi\fi\fi
   \qty{\fpeval{\result}}{\percent}%
}
\newcommand*\NumAndPercent[3][1]{\num{\fpeval{#2}}~(\ThePercent[#1]{#2}{#3})}

\usepackage{xspace}
\def\cran{\textsc{cran}\xspace}

\usepackage[inline]{enumitem}
\newlist{inlist}{enumerate*}{1}
\setlist[inlist]{itemjoin={{, }},itemjoin*={{, and }},label=(\(\roman*\)),mode=boxed}
\newlist{blocklist}{enumerate*}{1}
\setlist[blocklist]{itemjoin={{\space}},itemjoin*={{\space}},label=(\(\roman*\)),mode=boxed}
\newlist{orlist}{enumerate*}{1}
\setlist[orlist]{itemjoin={{, }},itemjoin*={{, or }},label=(\(\alph*\)),mode=boxed}

\definecolor{spaceorange}{HTML}{ED820E}

\usepackage{csquotes}
\usepackage[capitalize,nameinlink,noabbrev]{cleveref}

\let\T\texttt
\let\say\enquote

\definecolor{@soc}{HTML}{FC7768}
\definecolor{@cran}{HTML}{5B84B2}

\def\ColorExample#1{\tikz[baseline={(0,-.2\baselineskip)}]{\draw[draw=black,fill=#1,opacity=.875] (0,0) circle[radius=.25\baselineskip];}}
\robustify\ColorExample

\usepackage[most]{tcolorbox}
\newcounter{subfinding}
\def\findingborderwidth{2pt}
\newtcolorbox[auto counter]{finding}{
 enhanced,breakable,
 width=\linewidth,
 colframe=gray!5,
 top=3pt, bottom=3pt,
 right=3pt,
 boxsep=0pt,
 leftrule=0pt,
 borderline west={\findingborderwidth}{0pt}{gray},
 sharp corners,
 hyphenationfix,
 grow sidewards by=2mm,
 left=2mm,
 colback=gray!5,
 before upper={\textbf{Finding \thetcbcounter:}\setcounter{subfinding}{0}\space\ignorespaces},%
 extras first and middle={overlay={%
    \begin{scope}[shift={(frame.south west)}]
            \path[fill=gray] (0,1pt) -| ++(\findingborderwidth,-1pt)
            -- ++(-0.5*\findingborderwidth,-\findingborderwidth)
            -- ++(-0.5*\findingborderwidth,\findingborderwidth)
            -- cycle;
    \end{scope}
   }},%
   extras middle and last={overlay={%
      \begin{scope}[shift={(frame.north west)}]
            \path[fill=gray] (0,-1pt) -| ++(\findingborderwidth,1pt)
               -- ++(-0.5*\findingborderwidth,\findingborderwidth)
               -- ++(-0.5*\findingborderwidth,-\findingborderwidth)
               -- cycle;
      \end{scope}
   }}
}
\def\subfinding{\space\stepcounter{subfinding}\textit{\roman{subfinding})}\space}

\usepackage{subcaption}
\usepackage{fontawesome}

\usepackage{listings}

\lstdefinelanguage{lR}{
   language=R,
   alsoletter={.},
   deletekeywords={packages},
   literate={<<}{{<\null<}}2 {>>}{{>\null>}}2 {::}{{:\kern-1.5pt:}}2 {:::}{{:\kern-1.5pt:\kern-1.5pt:}}3
}
\DisableLigatures[<,>]{encoding=*}

\lstnewenvironment{R}[1][]
{%
    \lstset{%
        language=lR, aboveskip=4pt, belowskip=4pt, #1
    }%
}{}

\def\bR{\lstinline[language=lR]}
\newsavebox\CranBlob
\newsavebox\SocialBlob

\def\GetSignificanceWithBonferroni#1{%
      \edef\basenumberbonferroni{\fpeval{#1*\TheNumberOfTests}}%
      \edef\digits{\fpeval{ceil(ln(abs(\basenumberbonferroni))/ln(10))}}%
      \ensuremath{10^{\digits}}\relax
}

\def\CohensDToWord#1{%
   \edef\tempres{\fpeval{#1}pt}%
   \ifdim\tempres=0pt no effect\else
   \ifdim\tempres<0.2pt very small\else
   \ifdim\tempres<0.35pt small\else
   \ifdim\tempres<0.65pt medium\else
   \ifdim\tempres<1.6pt large\else
   \ifdim\tempres<2pt very large\else 
   huge\fi\fi\fi\fi\fi\fi
}

\def\SigEffMann{\stepcounter{numberoftests}\BasicSigEffMann}
\def\BasicSigEffMann#1#2{({\sisetup{tight-spacing}\ifnum#1=0\relax\def\currexp{-15}\else \def\currexp{#1}\fi\small\(p_{\scalebox{.65}{m}}\,{<}\,\GetSignificanceWithBonferroni{10^{\currexp}}\), \(d\,{=}\,\edef\absval{\fpeval{abs(#2)}pt}\ifdim\absval<0.01pt \sisetup{round-pad=false,round-precision=3}\fi\num{#2}\), \textit{\CohensDToWord{#2}}})}
\def\SigEffFisher{\stepcounter{numberoftests}\BaseSigEffFisher}
\def\BaseSigEffFisher#1#2{({\sisetup{tight-spacing}\small\(p_{\scalebox{.65}{f}}\,{<}\,\GetSignificanceWithBonferroni{#1}\), \(\phi\,{=}\,\sisetup{round-pad=false,round-precision=2}\edef\absval{\fpeval{abs(#2)}pt}\ifdim\absval<0.01pt \sisetup{round-pad=false,round-precision=3}\fi\ifdim#2pt<0pt \kern-2pt\fi\num{#2}\)})}

\newcounter{numberoftests}
\providecommand\TheNumberOfTests{-1}
\makeatletter
\AtEndDocument{\immediate\write\@auxout{\gdef\string\TheNumberOfTests{\number\value{numberoftests}}}}

\newread\fisherRes

\def\SigOutputFile{significance-tests.md}
\def\CleanTempFile{\immediate\write18{rm -f "\RTempFile"}}
\AtBeginDocument{\CleanTempFile}
\immediate\write18{rm -f "\SigOutputFile"}

\pgfqkeys{/msr-significance}{
   a with/.forward to        = {/msr-significance/research with},
   a without/.forward to     = {/msr-significance/research without},
   research with/.code       = \gdef\oResWith{#1},
   research without/.code    = \gdef\oResWithout{#1},
   b with/.forward to        = {/msr-significance/cran with},
   b without/.forward to     = {/msr-significance/cran without},
   cran with/.code           = \gdef\oCRANWith{#1},
   cran without/.code        = \gdef\oCRANWithout{#1},
   with/.code                = \gdef\oWith{#1},
   without/.code             = \gdef\oWithout{#1},
   name/.code                = \gdef\oName{#1},
   a name/.forward to        = {/msr-significance/research name},
   cran name/.code           = \gdef\oCRANName{#1},
   b name/.forward to        = {/msr-significance/cran name},
   research name/.code       = \gdef\oResName{#1},
   @defaults/.style          = {
      cran name              = CRAN,
      research name          = Research,
      research without       = \fpeval{\ProcessedSocialFiles-\oResWith},
      cran without           = \fpeval{\ProcessedCRANFiles-\oCRANWith},
      with                   = ??,
      name                   = \oWith,
      without                = not \oWith%
   }
}

\edef\hashtagsymbol{\string#}%
\makeatletter
\edef\backslashsymbol{\@backslashchar}%

\def\RTempFile{.fisheR.tmp}

\def\ExecuteR#1{\immediate\write18{%
R --vanilla --no-readline --quiet --no-echo -e '#1' > \RTempFile%
}}

\def\ReadResult#1{%
   \IfFileExists{#1}{%
      \immediate\openin\fisherRes=#1\relax
      \read\fisherRes to \result
      \immediate\closein\fisherRes
   }{\def\result{??}}%
   \typeout{=== RESULT: \result =======================}%
}

\let\get=\fpeval

\def\Cached#1#2{%
   \ifcsname cache:#1\endcsname
   \else
      #2
   \fi
}

\newwrite\cacheFile

\def\Cache#1#2{%
   \immediate\write\cacheFile{\string\expandafter\string\gdef\string\csname\space cache:#1\string\endcsname{#2}}%
   \expandafter\xdef\csname cache:#1\endcsname{#2}%
}
\def\GetCache#1=#2{%
   \ifcsname cache:#2\endcsname
      \edef#1{\csname cache:#2\endcsname}%
   \else
      \edef#1{??}%
   \fi
}

\AtBeginDocument{\input \jobname.cache\relax \immediate\openout\cacheFile=\jobname.cache\relax}
\AtEndDocument{\immediate\closeout\cacheFile}

\def\ExactFisheR#1{%
   \pgfqkeys{/msr-significance}{@defaults,#1}%
   \typeout{=== \oName: \oCRANName.\oWith: \oCRANWith, \oWithout: \oCRANWithout, \oResName.\oWith: \oResWith, \oWithout: \oResWithout =======================}%
   \Cached{\oName}{%
      \ExecuteR{v <- stats::fisher.test(matrix(c(\oResWith,\oResWithout,\oCRANWith,\oCRANWithout),nrow=2)); options(scipen=999); cat(v$ p.value)}%
      \typeout{CALCULATE}
      \ReadResult{\RTempFile}%
      \edef\fisherValue{\result}%
      \Cache{\oName}{\fisherValue}%
      \ExecuteR{phi <- psych::phi(matrix(c(\oResWith,\oResWithout,\oCRANWith,\oCRANWithout),nrow=2), digits=8); options(scipen=999); cat(phi)}%
      \ReadResult{\RTempFile}%
      \Cache{\oName-phi}{\result}%
      \CleanTempFile
   }%
   \def\n{^^J}%
   \GetCache\fisherValue={\oName}%
   \Cache{\oName}{\fisherValue}%
   \GetCache\phivalue={\oName-phi}%
   \Cache{\oName-phi}{\phivalue}%
   \immediate\write18{%
      echo '\n\hashtagsymbol\hashtagsymbol\space Significance for `\oName`\n\n | | \oResName | \oCRANName | |\n|:--|---:|---:|---:|\n | \oWith | \get\oResWith | \get\oCRANWith | \get{\oResWith+\oCRANWith} | \n | \oWithout | \get\oResWithout | \get\oCRANWithout | \get{\oResWithout+\oCRANWithout} | \n | | \get{\oResWith+\oResWithout} | \get{\oCRANWith+\oCRANWithout} | \get{\oResWith+\oResWithout+\oCRANWith+\oCRANWithout} |\n\n Using fisher'\backslashsymbol''s exact test: $ p < \fisherValue $ (phi-correlation coefficient: $ \phi = \phivalue $)\space (appears on page \thepage)\n\n' >> "\SigOutputFile"%
   }%
   \edef\toosmall{\fpeval{\fisherValue<10^(-15)?1:0}}%
   \ifnum\toosmall=1 \def\fisherValue{10^{-15}}\fi
   \SigEffFisher{\fisherValue}{\phivalue}%
}

\def\LoadFisher#1{%
   \GetCache\fisherValue={#1}%
   \GetCache\phivalue={#1-phi}%
   \edef\toosmall{\fpeval{\fisherValue<10^(-15)?1:0}}%
   \ifnum\toosmall=1 \def\fisherValue{10^{-15}}\fi
   \BaseSigEffFisher{\fisherValue}{\phivalue}%
}

\usepackage[super]{nth}
\usepackage{xfrac}
\def\ReviewLabel#1{\phantomsection\label{review-change:#1}}
\def\ReviewRef#1{\hyperref[review-change:#1]{page~\pageref*{review-change:#1} (\cref*{review-change:#1})}}

\begin{document}

\title{On the Anatomy of Real-World R Code for Static Analysis}

\author{Florian Sihler}
\email{florian.sihler@uni-ulm.de}
\orcid{0000-0001-7195-7801}

\affiliation{%
  \institution{Ulm University}
  \country{Germany}
}

\author{Lukas Pietzschmann}
\email{lukas.pietzschmann@uni-ulm.de}
\orcid{0009-0002-7803-6583}
\affiliation{%
  \institution{Ulm University}
  \country{Germany}
}

\author{Raphael Straub}
\email{raphael.straub@uni-ulm.de}
\orcid{0009-0007-8534-0053}
\affiliation{%
  \institution{Ulm University}
  \country{Germany}
}

\author{Matthias Tichy}
\email{matthias.tichy@uni-ulm.de}
\orcid{0000-0002-9067-3748}
\affiliation{%
  \institution{Ulm University}
  \country{Germany}
}

\author{Andor Diera}
\email{andor.diera@uni-ulm.de}
\orcid{0009-0001-3959-493X}
\affiliation{%
  \institution{Ulm University}
  \country{Germany}
}

\author{Abdelhalim Dahou}
\email{Abdelhalim.Dahou@gesis.org}
\orcid{0000-0001-8793-2465}
\affiliation{%
  \institution{GESIS - Institute for the Social Sciences}
  \country{Germany}
}
\renewcommand{\shortauthors}{Sihler et al.}

\def\blk#1{\kern1ex\textsc{#1}~\ignorespaces}
\begin{abstract}
   \blk{Context} The R~programming language has a huge and active community, especially in the area of statistical computing. Its interpreted nature allows for several interesting constructs, like the manipulation of functions at run-time, that hinder the static analysis of R~programs. At the same time, there is a lack of existing research regarding how these features, or even the R~language as a whole are used in practice.
   \blk{Objective} In this paper, we conduct a large-scale, static analysis of more than \num{50}~million lines of real-world R~programs and packages to identify their characteristics and the features that are actually used.
   Moreover, we compare the similarities and differences between the scripts of R~users and the implementations of package authors. We provide insights for static analysis tools like the \T{lintr} package as well as potential interpreter optimizations and uncover areas for future research.
   \blk{Method} We analyze \num{4230}~R~scripts submitted alongside publications and the sources of \num{19450}~\cran packages for over \num{350000}~R~files, collecting and summarizing quantitative information for features of interest.
   \blk{Results} We find a high frequency of name-based indexing operations, assignments, and loops, but a low frequency for most of R's reflective functions.
	Furthermore, we find neither testing functions nor many calls to R's foreign function interface~(FFI) in the publication submissions.
   \blk{Conclusion} R~scripts and package sources differ, for example, in their size, the way they include other packages, and their usage of R's reflective capabilities.
   We provide features that are used frequently and should be prioritized by static analysis tools, like operator assignments, function calls, and certain reflective functions like~\T{load}.
\end{abstract}



\received{17 November 2023}

\maketitle

\section{Introduction}
The R~programming language is primarily used in the area of data science. Its ecosystem of packages provides a vast set of predefined functions for statistical analysis~\cite{rcoreteam_intro_2023,smith2011r}.
Both, its large collection of libraries, as well as its mix of features from object-oriented and functional programming helped~R to grow popular in the statistics community.
Consequently, R's~user base primarily consists of people without a background in computer science~\cite{wickham_advanced_2019,goel2021we}. This motivates the creation of static analysis tools that help these users reuse existing, write new, and understand code better in general.
However, R's design as an interpreted and inherently dynamic language allows for a vast set of interesting features like the modification of functions at run-time, the evaluation of code from strings, and the lazy evaluation of function arguments~\cite{rcoreteam_language_2023,morandat2012evaluating,goel2021promises}.
These features hinder static analysis, as they require an analyzer to keep track of many potential run-time traces.
For example, to resolve the definition of a variable in the program or to retrieve the current implementation of a function.
This problem is underlined by a lack of static analysis tools for~R, with the tools using static analysis tending to be shallow. \ReviewLabel{additional-rw}\T{lintr}~\cite{lintr23}, the most popular linter for~R, relies mostly on XPath expressions, while other tools like \T{rstatic}~\cite{ulle2022source} and \T{ROSA}~\cite{sen2017rosa} ignore the (side-)effects of function calls.\par
Although existing research already investigated the dynamic usage of certain features at run-time~\cite{goel2019design,goel2021promises,goel2021we}, we are unaware of a static perspective on feature usage.
As R~allows a lot of problematic constructs, we want to find out which are often used in practice, hoping to support and encourage the development of more advanced static analyzers for R~programs.
This analysis is driven by the following three research questions:
\begin{enumerate}[label={\textsc{rq}\arabic*:},ref={RQ\arabic*},leftmargin=*,nosep]
	\item \label{rq:main}\textit{What are common and lesser used features in R programs?} Our data shows that R~users tend to frequently use name-based indexing, \T{for} loops, and \T{if} structures.
	\item \label{rq:diff}\textit{How do the used features differ between code written by researchers to analyze data and package authors?} We find that research scripts completely lack any use of the FFI, as well as testing functions while using \T{for} loops more often.
	\item \label{rq:cons}\textit{What are the insights gained for static analysis tools?} Albeit most reflective features are used rarely, we find that static analysis tools should nevertheless consider a subset of them like the \T{load}~function.
\end{enumerate}

To address these research questions, we collect R~files from two sources:
\begin{inlist}
    \item R~scripts, written to analyze a specific dataset using statistical tests and plots
    \item R~sources, written to be reused by others in the form of packages
\end{inlist}.
For these datasets, we use the AST of R~programs combined with the dataflow information to extract quantitative information about the usage of features in the source code. Furthermore, we manually inspect a subset of the results to uncover the purpose and reason for their usage.

Since our information extraction is intentionally limited to static analysis, we cannot provide any information on dynamic properties, such as determining the number of calls during runtime.

\ReviewLabel{motivation-intro}We identify the following practical applications of our work:
\begin{blocklist}
	\item \textit{Static analysis tools:} Understanding how R's language constructs are used in practice enables developers of static analysis tools to prioritize features to support, for example by focusing on functions like \T{load} and \T{attach} when handling reflective functions, and to speed up the analysis with optimistic assumptions.
	\item \textit{Interpreter optimization:} R~interpreter developers can add special cases for common constructs or establish optimistic assumptions based on the knowledge that lesser-used features, like the redefinition of primitive functions, do not occur. However, even though static data can offer valuable insights into the location and sequencing of events, the interpretations must be taken with care, as statically extracted information does not necessarily reflect the dynamic behavior of a program.
	\item \textit{Teaching:} Our results can be used to identify features that are used rarely and therefore might be less relevant to teaching, as well as common smells that should be addressed.
	\item \textit{R language development:} Similarly, our data can be used to make informed decisions about the deprecation of language features.
	\item \textit{Future research:} Our data collection can be used to uncover future research areas. For example, it suggests the effectiveness of error-mining techniques for R~programs and the potential for abstract interpretation.
\end{blocklist}

Our findings mainly benefit developers of static analyzers and other tools that assist people in writing better R~code, which, by proxy, benefits the R~community as a whole.
Altogether, we provide the following contributions:
\begin{enumerate}[nosep,wide=0pt]
    \item An automated approach to extract quantitative information about the usage of features in R~programs.
    \item A reusable dataset of over \num{50}~million lines of R~code.
    \item Usage statistics to help focus on supporting the most widely used features in static analyzers like linters.
\end{enumerate}

In \cref{sec:related-work}, we first provide an overview of related work in the large-scale analysis of programming languages. Afterward, we describe our methodology in \cref{sec:methodology} and present our findings in \cref{sec:findings}. We then discuss our results and provide actionable results in \cref{sec:discussion}, followed by a discussion of our threats to validity in \cref{sec:threats}. Finally, we conclude our work in \cref{sec:conclusion}. For more information on the artifacts provided, refer to \cref{sec:data-availability}.

\section{Related Work}\label{sec:related-work}
\textit{Language Usage Studies.} One of the first researchers to conduct an empirical study on the usage of language constructs was \citeauthor{knuth_empirical_1971} in~\citeyear{knuth_empirical_1971}~\cite{knuth_empirical_1971}.
He leveraged static analysis to compile a distribution
of statement types and dynamic analysis for statement
executions on a total of 440~Fortran programs, intending to provide more
informed optimizations for the Fortran compiler. The paper led many
researchers to investigate different languages.
In~\citeyear{dyer_mining_2014}, \citeauthor{dyer_mining_2014} statically analyzed the~AST of over
9~million Java files~\cite{dyer_mining_2014,dyer_boa_2013}, focusing on the
adoption of new features over three main language revisions.
Three years later, \citeauthor{qiu_understanding_2017} conducted a similar AST-based analysis
to identify the usage of Java's syntactic rules~\cite{qiu_understanding_2017}.
\textit{R's Capabilities.}
While an informal definition
of the R~programming language and its features exists~\cite{rcoreteam_definition_2023},
it does not provide a complete picture of R's capabilities and neither information on their usage in practice
nor their semantics.
\citeauthor{wickham_advanced_2019} focuses on approachable explanations for various language features instead of defining the language itself~\cite{wickham_advanced_2019}.
While mentioning the focus on \say{useful} features, he does not give any insight into
what qualifies as useful.
In the absence of a formal language definition and with only incomplete
documentation available~\cite{morandat2012evaluating}, \citeauthor{burns_inferno_2011}
tries to assist users in navigating various pitfalls, peculiar design decisions, and
language-related challenges~\cite{burns_inferno_2011}. While providing helpful insight
into R's mechanics, he does not propose a solution to the underlying problem by way of a
formal definition. To our knowledge, \citeauthor{morandat2012evaluating} are the first
to address the aforementioned foundational issues for R~(version 2.12.1,~\cite{morandat2012evaluating}).

\textit{Analysis of R Programs.} Besides formalizing~R, \citeauthor{morandat2012evaluating} introduce TraceR. A software suite capable of
static and especially dynamic
code analysis using a modified R virtual machine~\cite{morandat2012evaluating}. In their work, they evaluate the current implementation of the interpreter by collecting information on memory and time complexity to
investigate the actual dynamic usage of several features of the R~programming language in a total of \num{3.9}~million lines of code.
This analysis provides valuable insights into both the theoretical foundations and real-world applications of R's features.
Our work improves on the results of \citeauthor{morandat2012evaluating} by \begin{inlist}
	\item using a newer version of the R interpreter (4.3.1 instead of~2.12.1)
	\item analyzing a much larger set of code
	\item covering a broader set of language features
	\item manually investigating the results
\end{inlist}. \ReviewLabel{jan-vitek-sep}Moreover, we differ from their approach, as we analyze the code from a static perspective. Results from static and dynamic analyses can differ as a single call site can cause many invocations.
Other publications study specific language
features, like laziness~\cite{goel2019design}, promises~\cite{goel2021promises}, and
\texttt{eval}~\cite{goel2021we}.
Our work differs in that we cover a much broader set of R~features.
Due to the low usage of lazy evaluation and its substantial performance impact and
memory usage~\cite{goel2019design}, \citeauthor{goel2021promises} propose a strategy to
evolve R into a strict language~\cite{goel2021promises}.

\begin{figure*}[t]
    \centering
    \newsavebox\FirstProcessingStep
    \savebox\FirstProcessingStep{\sffamily\begin{tikzpicture}[xscale=.9,yscale=.9]
       \node[start] (start) at(0,0) {};
       \node[block,right=2.5mm] (parse) at(start.east) {~~Parse~~};
       \node[block,right=6mm] (df) at(parse.east) {~~Dataflow~~};
       \node[block,right=6mm] (extract) at(df.east) {~~Extract~~};
       \node[start,right=2.5mm] (end) at(extract.east) {};
       \draw (end) circle[radius=1.33ex];

       \pgfinterruptboundingbox
       \node[above=-1.5pt,T] at(parse.north) {\strut\num{123} errors};
       \node[above=-1.5pt,T] at(df.north) {\strut\num{3} heap, \num{21} form. errors};

       \node[below=-1.5pt,T] at(parse.south) {\strut\num{1156} errors};
       \node[below=-1.5pt,T] at(df.south) {\strut\num{10} heap, \num{199} form. errors};

       \endpgfinterruptboundingbox

       \draw[flow] (start) -- (parse);
       \draw[flow] (parse) -- (df);
       \draw[flow] (df) -- (extract);
       \draw[flow] (extract) -- (end.west);
    \end{tikzpicture}}
\resizebox{.91\linewidth}!{\sffamily\begin{tikzpicture}[xscale=.9,yscale=.9]

   \node[artifact=@soc] (ssoc-main) at(0,0) {\strut Research-Sources\strut};
   \node[artifact=@cran, below=5mm] (cran-main) at(ssoc-main.south) {\strut CRAN-Packages\strut};

   \coordinate[right=.5cm] (@) at(current bounding box.east);
   \node[right,draw=gray,rounded corners=2pt,inner sep=1.1em] (first-step) at (@) {\usebox\FirstProcessingStep};

   \draw[flow] ([yshift=2mm]ssoc-main.east) node[above right,T] {\num{4230} files} -| ([xshift=1.25em+.5em]first-step.north west);
   \draw[flow] ([yshift=-2mm]cran-main.east) coordinate (@cran-marker) node[below right,T] {\num{358989} files} -|  ([xshift=1.25em+.5em]first-step.south west);

   \node[artifact,right=.5cm] (file-based-results) at (first-step.east|-ssoc-main.east) {File-Based\\Results};

   \draw[flow] ([xshift=-1.25em-.5em]first-step.north east) |- ([yshift=2mm]file-based-results.west) coordinate[pos=.5] (@soc-numnode);
   \draw[flow] ([xshift=-1.25em-.5em]first-step.south east) |- ([xshift=.5em]@cran-marker-|first-step.south east) coordinate[pos=.5] (@cran-numnode) |- ([yshift=-2mm]file-based-results.west);

   \node[block] (summarize) at (file-based-results|-cran-main) {Summarize};
   \draw[flow] (file-based-results) -- (summarize);

   \node[block,right=.5cm] (vis) at(current bounding box.east) {\rotatebox{90}{\parbox{2.5cm}{\centering Visualization and\\Manual Investigation}}};

   \draw[flow] (summarize.east) -- ++(3mm,0) |- (vis.west);

   \node[above left,xshift=9mm,T] at(@soc-numnode.north east) {\num{\ProcessedSocialFiles} files};
   \node[below left,xshift=9mm,T] at(@cran-numnode.south east) {\num{\ProcessedCRANDefaultFiles} files (\num{\ProcessedCRANExampleFiles} examples, \num{\ProcessedCRANTestFiles} tests)};

\end{tikzpicture}}
    \caption[The Extraction Workflow.]{The Extraction Workflow. Of the \num{\SocialFiles}~research files, \NumAndPercent{123}{\SocialFiles} failed to parse, \NumAndPercent{3}{\SocialFiles} caused a heap memory overflow, and \NumAndPercent{21}{\SocialFiles} either contained encoding errors or features that our framework could not handle. Similarly, of the~\num{\CRANFiles} \cran files, \NumAndPercent{1156}{\CRANFiles} failed to parse, \NumAndPercent[2]{10}{\CRANFiles} caused a heap memory overflow, and \NumAndPercent{199}{\CRANFiles} failed to process (cf. \cref{sec:result-error}).}
    \label{fig:main-workflow}
\end{figure*}
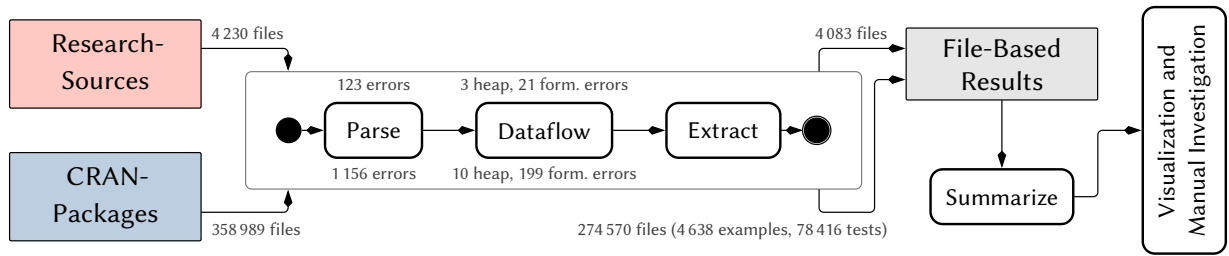

\section{Methodology}\label{sec:methodology}
We base our procedure on the systematic process described by \citeauthor{vidoni_systematic_2022}~\cite{vidoni_systematic_2022} which provides seven guidelines split over three main phases: \textit{planning}~(\ref{sec:meth-planning}), \textit{execution}~(\ref{sec:meth-execution}), and \textit{results}~(\ref{sec:meth-results}).
Furthermore, we follow the ACM SIGPLAN~\cite{acmSigplanGuidelines} and SIGSOFT Empirical Standards~\cite{acmSigsoftGuidelines} and consider ethical implications~\cite{gold_ethics_2021}.

\subsection{Planning}\label{sec:meth-planning}
This section outlines the research objectives and sources used~\cite{vidoni_systematic_2022}.

\subsubsection{Research Objectives} We want to identify the features that R~programmers use in their code and the way these features differ between
code written by researchers for analyzing data and package authors, as expressed by our research questions.
From this, we do not only want to obtain a better understanding of the R~programming language and its usage but also provide a basis for future research on static analysis of R~programs.

\subsubsection{Sources}\label{ssec:meth-sources} We collect R sources in two categories: supplements to publications and sources of R~packages.

To retrieve R~scripts used as supplements of publications, we employ three commonly used open-source platforms: \begin{inlist}
    \item the \textit{Journal of Statistical Software}~(\num{22}~projects)
    \item \textit{Figshare}~(\num{49}~projects)
    \item \textit{Zenodo}~(\num{1534}~projects)
\end{inlist}.\footnote{Found at \url{https://www.jstatsoft.org}, \url{https://figshare.com}, and \url{https://zenodo.org}.} 
We scraped these data sources in May~2023 using the \enquote{Instant Data Scraper} chrome extension and a custom Python script to download and extract the source files.
For the subsequent analysis, we filtered the sources for R~code by only keeping files ending in either~\say{\T{.r}} or~\say{\T{.R}}, which results in \num{4230}~files. To uniformly process and parse all files, we chose to exclude other well-known formats containing R~code, like RMarkdown~\cite{rmarkdown}.
We refer to these sources as the \say{research dataset}.

To obtain the code of packages, we selected the Comprehensive R~Archive Network~(\cran) as the data source, as it is the primary place to publish R~packages. We downloaded the raw source code of all \num{19450}~packages available on May~5,~2023.

Again, we retain only files ending in either~\say{\T{.r}} or~\say{\T{.R}}, which results in \num{358989}~files.
We refer to it as the \say{\cran dataset}.

\ReviewLabel{other-repositories}We decided against including sources like \textit{GitHub} or \textit{GitLab} because they contain a mixture of R~package code and single-file scripts for various purposes, which would require additional filtering steps.
We argue, that because of the large size of the \cran dataset, this does not impact our findings.

\subsubsection{Used Versions}
We conducted every automated step of our process on the same system 
using R~version~\textit{4.3.1},
\say{Beagle Scouts}. For the parsing, we additionally relied on the \T{xmlparsedata} package version~\textit{1.0.5}~\cite{xmlparsedata21}. Our extractor script is written in TypeScript and runs on Node.js version~\textit{18.17.1} (see \cref{sec:data-availability}).

\subsection{Execution}\label{sec:meth-execution}
For the data and information extraction, we use a three-step workflow, summarized in \cref{fig:main-workflow}.
After collecting both datasets (cf.~\cref{ssec:meth-sources}), we analyze each file completely automatically.

We start by using the \T{parse} function of the R~interpreter~\cite{rcoreteam_language_2023a} and the \T{xml\_\-parse\_\-data} function of the \T{xmlparsedata} package~\cite{xmlparsedata21} to obtain the AST in XML format. The procedure is similar to that of the well-known R~languageserver~\cite{languageserver23} and the \T{lintr} package~\cite{lintr23}.

Given the~AST of the parsing step, we extract dataflow information~\cite{khedker2017data} by folding the tree, keeping track of variable references, definitions, and function calls to produce a single static dataflow graph for the whole file. Even though this extraction can not handle several of R's reflective capabilities~(cf. \cref{sec:threats}), it allows us to identify targets of called functions, variable re-definitions, obviously dead code, and more.

For our open research objective of identifying the features R~users utilize, we extract ten~different characteristics that we consider indicative, based on consulting existing sources like the R~Language Definition~\cite{wickham_advanced_2019,rcoreteam_definition_2023,morandat2012evaluating}: assignments, comments, conditionals, ways of data access, function definitions, function calls, loops, used packages, values, and variable uses.
For each characteristic, we use either \begin{orlist}
    \item a set of XPath expressions, based on those used by the \T{lintr} package~\cite{lintr23} or the R~language server~\cite{languageserver23}
    \item a visitor on the AST
\end{orlist}. \ReviewLabel{dataflow}We limit our utilization of the dataflow information to tracking the targets of function calls as well as the identification of (re-)defined variables, as our graph makes identifying them trivial.

\subsection{Results}\label{sec:meth-results}
This final phase is responsible for the summarization, analysis, and presentation of the extracted data.
In the first step, we aggregate the file-based results from the extraction. For the \cran-package dataset, we additionally separate the results into three categories: \begin{inlist}
    \item \textit{examples}: if the file path contains \say{example}
    \item \textit{tests}: if the file path contains a test prefix (excluding the package name to compensate for test packages like \T{testthat})
    \item \textit{default}: in all other cases
\end{inlist}.
Although we usually treat all of these categories as one, some features are used differently based on the general purpose of the file (e.g., testing functions in test files). We mention these differences whenever they are of interest.

\ReviewLabel{significance-methodology}To back up our claims, we calculate the significance using Fisher's exact test~\cite{fisher_interpretation_1922} and the Mann-Whitney test~\cite{mann_test_1947}.
However, due to the large sample size, the reported significance values are very low (i.e., represent a very high significance). We adjust the p-values using the Bonferroni correction with \(k = \TheNumberOfTests\), based on the number of tests~\cite{cabin2000bonferroni}. Additionally, we use the phi-coefficient~\cite{cramer1946mathematical} for the Fisher's exact test and Cohen's~\(d\) with a~\qty{95}{\percent} confidence interval for the Mann-Whitney test~\cite{cohen_statistical_1988} to measure the effect size. With \enquote{\SigEffMann{-5}{0.25}} we report a significance of~\(p_m < 0.001\) using Mann-Whitney, the Bonferroni correction already applied, and an effect size of~\(d = 0.25\)~--- which, using the interpretation of \citeauthor{sawilowsky2009new}~\cite{sawilowsky2009new}, represents a \textit{\CohensDToWord{0.25}} effect.
\addtocounter{numberoftests}{-1}%

Even though usage frequencies of features can already reveal a lot, they do not reveal their \textit{purpose} in the code (i.e., \textit{why} they are used). Therefore, we sometimes provide anecdotal evidence for the purpose, based on a manual investigation of a subset of the results.

We use~R and \LaTeX{} to semi-automatically visualize the results for each feature in \cref{sec:findings}, based on the summarized data.

\section{Findings}\label{sec:findings}
This section is split into eight parts, each focusing on a different aspect of our extraction: the files we were able to process~(\ref{sec:result-error}), the differences in file size~(\ref{sec:result-meta}), the ways R~users assign data~(\ref{sec:result-assignments}), write conditions~(\ref{sec:result-conditionals}), loops~(\ref{sec:result-loops}), function definitions~(\ref{sec:result-fun-def}), the function R~users call~(\ref{sec:result-fun-call}), and the packages they use~(\ref{sec:result-packages}).

\begin{figure}
    \centering
    \includegraphics[width=\linewidth]{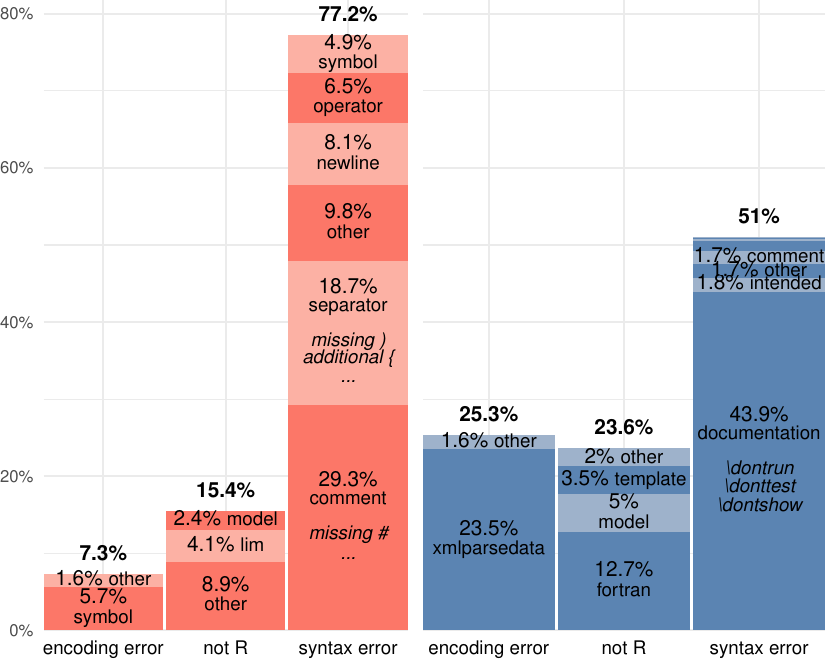}
	\caption{Causes of parsing errors. The left chart categorizes all \num{123}~parsing errors of the \ColorExample{@soc}~research dataset, the right does the same for the \num{1156}~failed files of the \ColorExample{@cran}~\cran dataset.}
    \label{fig:parse-errors}
\end{figure}

\subsection{Processing Failures}\label{sec:result-error}
We were unable to analyze \NumAndPercent{123+3+21}{\SocialFiles} of \num{\SocialFiles} files from the research sources and \NumAndPercent{1156+10+199}{\CRANFiles} of the~\num{\CRANFiles} files of the \cran-package dataset respectively. Manually investigating all failures revealed interesting findings regarding the syntax errors these files contained.

\subsubsection{Parse}\label{ssec:find-parse-failures}

\cref{fig:parse-errors} presents the manually classified reasons for which the respective files failed.\footnote{All errors indicate the \textit{first} error at which the R~parser halted the execution.}
In the research sources, \qty{15.4}{\percent} of the failed files were not R~code to begin with (wrongly suffixed with either \T{.r} or \T{.R}), but, for example, input files for linear inverse models~(lim).
\qty{7.3}{\percent}~contained wrongly encoded characters, like the \say{Fullwidth Equals Sign} Unicode character instead of the similar-looking \say{Equals Sign}.
However, the majority of files~(\qty{77.2}\percent) contained normal R~code with conventional syntax errors, like missing closing parenthesis or missing comment symbols in front of explanatory remarks or banners.

Similarly,~\NumAndPercent{1156}{\CRANFiles} of \num{\CRANFiles} files in the \cran dataset failed to parse of which only \qty{1.7}{\percent} appeared to be intended errors used as part of a test.
{\let\bs\textbackslash Most of these files~(\qty{43.9}{\percent}) contained one of the three (\LaTeX-)commands \T{\bs dontrun}, \T{\bs donttest}, or \T{\bs dontshow} intended for R~documentation files.}

We exclude all of these files from the following steps.

\subsubsection{Dataflow}\label{ssec:find-dataflow-failures}
In total, \num{12}~files exceeded our heap memory which was limited to \qty{4}{\giga\byte}. \num{220}~files contained structures that our dataflow extraction could not handle, with the two largest contributors being: \begin{inlist}
    \item \NumAndPercent{21+144}{220} Unicode characters that caused the \T{xml2js} library we use internally, to return an AST with an unexpected structure
    \item \NumAndPercent{45}{220} raw character strings in a form like \T{r"(...)"} which we were unable to handle at the time (they appeared exclusively in the \cran dataset)
\end{inlist}.

Similarly to the files that failed to parse, we exclude these files from the following steps.
Ultimately, we were able to process \num{4083}~fi\-les of the research and \num{357624}~fi\-les of the \cran-dataset.

\begin{finding}
\NumAndPercent{95}{\SocialFiles} of all research files fail to parse with R's parser due
	to syntax errors. \NumAndPercent{9}{\SocialFiles} contained unsupported characters. For \cran-packages, most of the \num{1156}~parse errors stem from documentation commands
in the code~(\qty{43.9}\percent), problematically encoded files~(\qty{23.5}\percent), and Fortran code with~\T{.r} or~\T{.R} file-endings~(\qty{12.7}{\percent}).
\end{finding}

\subsection{Metadata}\label{sec:result-meta}

Comparing the file sizes in the two datasets shows the median research script to be around~\TheFrac{6744}{2186} times the size of the median \cran file \SigEffMann{-15}{0.490757311751383}, with~\ThePercentIncrease{39.02}{32.09} longer lines \SigEffMann{-15}{0.00354923437622407}.
Manually investigating these files suggests that this is due to research scripts often containing the complete code for a submission while R~packages split their code into several files.

However, there are big outliers, with the largest \cran file being~\TheFracIncrease[1]{4245886}{2186} times the size of the median file (compared to a factor of~\TheFracIncrease[1]{319818}{6744} for the largest file in the research dataset).\footnote{The respective file is part of the \T{happign} package and contains a single test that expects a large message of geographical coordinates.}
For \cran-packages, test- and example-files tend to be only half the size of \enquote{normal} package code.

\begin{finding}
Publication supplements tend to be three times the size of the median \cran-package source with~\ThePercentIncrease{39.02}{32.09} longer lines. The 99\textsuperscript{th} percentile of files does not exceed \num{49226}~characters.
\end{finding}

\begin{figure}
    \centering
    \includegraphics[width=\linewidth]{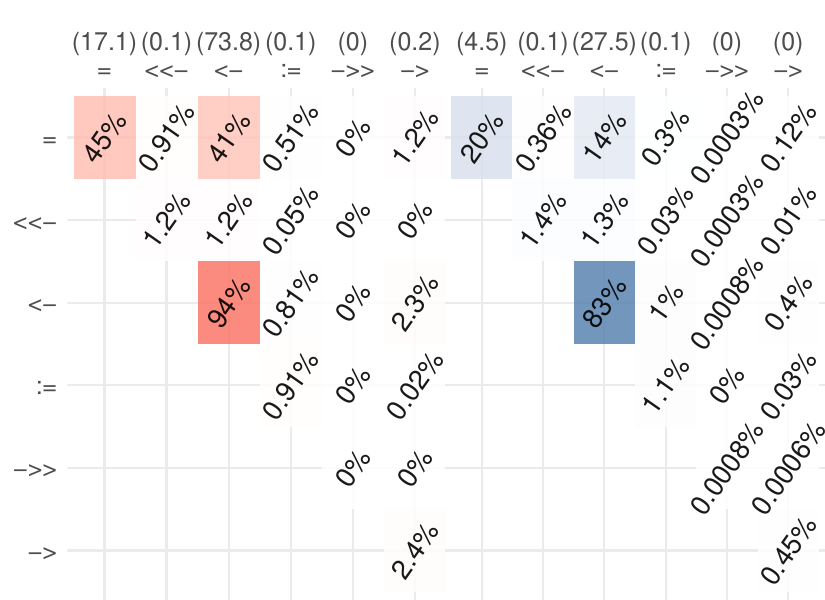}
	\caption{Heatmap of the usage of different assignment operator combinations. Left, the results for the \ColorExample{@soc}~research and right, the results for the \ColorExample{@cran}~\cran dataset.  For example,~\qty{94}{\percent} of all research~files use~\T{<-} for assignments. \qty{1.2}{\percent} use~\T{<<-} and~\T{<-}. As the numbers above indicate the average usage per file, no row or column needs to sum up to \qty{100}{\percent}.}
    \label{fig:assignments}
\end{figure}

\subsection{Assignments and Data Access}\label{sec:result-assignments}
For the way that R~users assign data, we separate three different perspectives: the operators and functions they use~(\ref{ssec:a-used},\,\ref{ssec:a-used-fun}), the number of re-definitions~(\ref{ssec:a-redef}), and the assigned values~(\ref{ssec:a-assigned}). Furthermore, we analyze the way R~users access data~(\ref{ssec:a-access}).

\subsubsection{Used Operators}\label{ssec:a-used}
R provides six assignment operators~\cite{rcoreteam_language_2023}: the left assignment~\T{<-}, which assigns its right-hand side~(RHS) to the left-hand side~(LHS) in the current environment similar to the assignment in other programming languages. The right assignment~\T{->} which works similarly, but with the roles of RHS and LHS swapped, the super assignments~\T{<<-} and~\T{->>} which modify existing variables in a parent environment~\cite[Chp.~7]{wickham_advanced_2019}, as well as the equal assignments~\T{=} and \T{:=}, which work similar to~\T{<-}.

\cref{fig:assignments} presents the frequencies of assignments used. Each cell represents the percentage of files in the dataset using the respective combination, which shows a clear dominance of~\T{<-}. This aligns with this operator being favored over~\T{=} by the \textit{tidyverse} style guide~\cite{tidyverseStyle23}.
However, the discouraged \say{\T{=}}-assign appears second-most~--- more than twice as often in research scripts as in \cran-package code~--- with a relatively large number of files (\qty{41}{\percent} and~\qty{14}{\percent}) mixing both aforementioned operators.
\ReviewLabel{operator-usage}As these operators have the same semantics when used as an assignment, there is no apparent reason to mix them.
Interesting is the relatively frequent usage of~\T{->} in the research dataset, often at the end of pipes:
\begin{R}
x %>% f %>% g -> result
\end{R}
However, this usage is discouraged by Google's fork of the \textit{tidyverse} style guide~\cite{googleRStyleGuide19}.
R's super assignments are rarely used, with~\T{<<-} clearly being favored over~\T{->>}, and a slight dominance in the \cran dataset. For example, to provide stateful functions that, when called again remember their previous state~\cite[Sec. 10.2]{wickham_advanced_2019}.

\subsubsection{Used Assignment Functions}\label{ssec:a-used-fun}
Besides the six operators,~R provides several functions to assign values, like \T{assign}, \T{assign\-In\-Namespace}, and \T{setGeneric}, which allow to assign values or functions to variables given as a string, potentially in a different environment~\cite{wickham_advanced_2019}. Together, these functions contribute to
just \NumAndPercent[2]{264+1}{\AllCallsSocial} of all calls in the research sources (\ThePercent{59+1}{\ProcessedSocialFiles} of files), compared to \NumAndPercent[2]{18174+59+3761+41294+12255}{\AllCallsCRAN} of calls in the \cran sources~(\ThePercent{4868+35+3214+7106+844}{\ProcessedCRANFiles} of files), with \T{setMethod} being the most common. In general, we can see that such assignment functions are more common in \cran files \SigEffMann{-12}{-0.0559435536726946}.

R's feature of preventing variables from redefinitions with \T{lock\-Environment} and \T{lock\-Binding}, is essentially unused, appearing only \num{2}~times in the research and \num{\fpeval{41+96}} times in the \cran dataset.

\begin{figure}
    \centering
    \includegraphics[width=\linewidth]{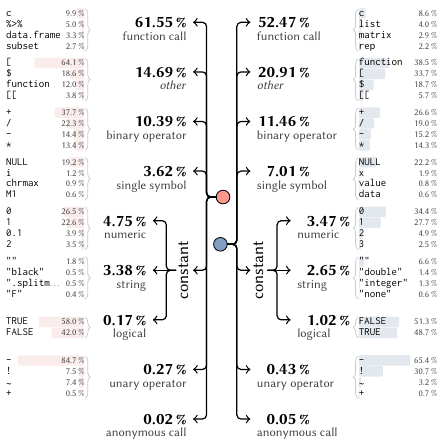}
	\caption{The distribution of different kinds of source expressions for assignments separated by the token type of the root node. Left, the \ColorExample{@soc}~research dataset and right the \ColorExample{@cran}~\cran dataset. For example,~\qty{10.39}{\percent} of all assignments in the research dataset assigned the result of a binary operation. Of those~\qty{10.39}{\percent} binary operations,~\qty{37.7}{\percent} are an addition with~\T{+}.}
    \label{fig:assigned-types}
\end{figure}
\begin{figure*}
    \centering
    \hspace*{-5pt}\begin{subfigure}[t]{\dimexpr.5\linewidth-5pt}
        \includegraphics[width=\linewidth,page=1]{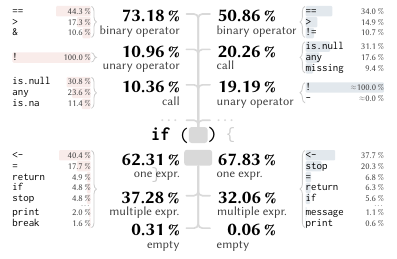}\vspace*{-.3\baselineskip}%
        \caption{\T{if}-constructs without an \T{else}-block. On average, \TheFrac{8349}{\ProcessedSocialFiles} per \ColorExample{@soc}~research and \TheFrac{1952793}{\ProcessedCRANFiles} per \ColorExample{@cran}~\cran file. Body-expressions are joined for \T{if} and \T{if-else}.\hfill\strut}
    \end{subfigure}\hfill
    \begin{subfigure}[t]{\dimexpr.5\linewidth-5pt}
        \includegraphics[width=\linewidth,page=2]{plots/if-types-compact.pdf}\vspace*{-.3\baselineskip}%
        \caption{\T{if}-constructs \hfill with \hfill an \hfill\T{else}-block.\hfill \TheFrac{4692}{\ProcessedSocialFiles} \hfill per \hfill \ColorExample{@soc}~research \hfill and \hfill \TheFrac{938115}{\ProcessedCRANFiles}\kern6.4pt\\per \ColorExample{@cran}~\cran file.\hfill\strut}
    \end{subfigure}\hspace*{-5pt}\vspace*{-.3\baselineskip}
    \caption{The distribution of expression types used in~\T{if} and~\T{if}-\T{else} constructs. Expressions in the condition are presented at the top and expressions in the body at the bottom, listing all three variants of an empty body, a body with a single or multiple expressions. The structure is similar to that shown in \cref{fig:assigned-types}.}
    \label{fig:if-types}
\end{figure*}

\subsubsection{Re-Definitions}\label{ssec:a-redef}
Not every file in the datasets uses the assignment operators. \NumAndPercent{4006}{\ProcessedSocialFiles} of the files from the research dataset and only \NumAndPercent{317449}{\ProcessedCRANFiles} of \cran files ever define a variable \ExactFisheR{
    research with    =                 4006,
    cran with        =               317449,
    with             =    {define variable}%
}.\footnote{Of the remaining files, \NumAndPercent{49}{\ProcessedSocialFiles-4006} and \NumAndPercent{244}{\ProcessedSocialFiles-317449} files are empty. The others either use other functions to define variables or only call other functions. See \cref{ssec:c-reflective}.}
Of those, \ThePercent{3605}{4006} research and only~\ThePercent{231414}{317449} \cran files ever redefine at least one variable explicitly \ExactFisheR{
    research with    =                 3605,
    research without =                 4006-3605,
    cran with        =               231414,
    cran without     =               317449-231414,
    with             =    {redefine variable}%
}. This excludes implicit \say{re-definitions} of parameters when calling a function multiple times and re-definitions of loop variables.

\subsubsection{Assigned Values}\label{ssec:a-assigned}

Considering the assigned values as shown in \cref{fig:assigned-types}, most assignments (\qty{61.55}{\percent} and~\qty{52.47}{\percent}) simply bind the result of a single function call, with the vector-construction function~\T{c} being the most-called in both datasets.
Less frequent is the direct assignment of constants~(\qty{\fpeval{4.75206057583509+3.37843272896926+0.172407321010098}}{\percent} and~\qty{\fpeval{3.47254785879645+2.6499876989788+1.01680328189507}}{\percent}) or single symbols (e.g., to alias a value).

Interestingly, we found several anonymous calls~--- functions that are defined or returned from other functions and called directly~--- in~\num{68} and~\num{5231} of cases in the research and \cran sources respectively, showing no clear dominance in either of the datasets \ExactFisheR{
    research with    =                 68,
    research without = \AllCallsSocial-68,
    cran with        =               5231,
    cran without     = \AllCallsCRAN-5231,
    with             =  {anonymous calls}%
}. Manually investigating the uses suggests two exclusive use cases in research sources (i.e., the only two found): 
\begin{inlist}
    \item color palette functions like \T{color\-Ramp\-Palette} which return a function that can interpolate a palette of~\(n\) colors
    \item the use of the \T{stats::ecdf} function, which returns an empirical cumulative distribution function, which may be called directly
\end{inlist}.

\subsubsection{Data Access}\label{ssec:a-access}
R provides several operators to access vectors, data-frames, and other structures~\cite{rcoreteam_definition_2023,wickham_advanced_2019}.
The single bracket access operator~\T{[}, which can be combined with a predicate to select elements,
is used on average \TheFrac{195016}{\ProcessedSocialFiles}~times per research and \TheFrac{4050460+15892+301559}{\ProcessedCRANFiles}~times per \cran file, with~\T{[[} accounting on average \TheFrac{19124}{\ProcessedSocialFiles}~and \TheFrac{1036736+4081+127707}{\ProcessedCRANFiles}~uses per file.
Interestingly, the access by name with~\T{\$}, which can only be used with constant strings (simplifying the identification of the accessed field significantly), dominates the index-access in both datasets, accounting to \TheFrac{271414}{\ProcessedSocialFiles}~uses per research and \TheFrac{3955133+25980+663678}{\ProcessedCRANFiles}~uses per \cran file. It accounts for \ThePercent{271414}{271414+19124+195016} and \ThePercent{3955133+25980+663678}{3955133+25980+663678+1036736+4050460+4081+15892+127707+301559} of all basic operator access operations with \T{[}, \T{[[}, and \T{\$}, revealing a slight trend in the research files \ExactFisheR{
    research with    =                 271414,
    research without =      19124+195016,
    cran with        =   3955133+25980+663678,
    cran without     =  1036736+4050460+4081+15892+127707+301559,
    with             =   {dollar access}%
}.

Besides simply using the name of a variable to reference it,\footnote{Which occurs on average~\TheFrac[1]{1297347}{\ProcessedSocialFiles} times per research and~\TheFrac{33069948}{\ProcessedCRANFiles} times per \cran file.} R provides functions like \T{get}, the by far most common, which searches for one or more objects by name. They amount for merely~\ThePercent{388+5}{\AllCallsSocial} of all calls in the research and~\ThePercent{26590+3844+41+415+800+31}{\AllCallsCRAN} in the \cran dataset.

\begin{finding}
    \subfinding Both datasets tend to assign values using the~\T{<-} operator recommended by the \textit{tidyverse} style guide. Yet, several research files still use~\T{=} for assignments, with~\qty{41}{\percent} even mixing~\T{<-} and~\T{=} for no apparent reason.
    Super assignments are rarely used.
    \subfinding Most assignments directly assign the results of function calls. Assigning new functions is slightly more common in the \cran dataset \ExactFisheR{
        research with    =                 6589,
        research without =          372954-6589,
        cran with        =               925018,
        cran without     =      11503110-925018,
        with             =    {assign function with op}%
    }.
    \subfinding \ThePercent{3605}{4006} of research and \ThePercent{231414}{317449} of \cran source files that define a value using one of the assignment operators ever redefine it (explicitly).
    \subfinding Assignment functions like \T{assign} and retrieval functions like~\T{get} are rarely used in research sources.
    \subfinding Index-Access by name is the most common. Research files, in general, contain more than three times as many index accesses as \cran files (which correlates with their size difference).
\end{finding}

\subsection{Conditionals}\label{sec:result-conditionals}
R~offers~\T{if} and~\T{if}-\T{else} constructs to conditionally execute code. However, there are several functions (e.g., \T{ifelse} and \T{switch}) that allow the alteration of control flow due to R's lazy evaluation of function arguments~\cite{wickham_advanced_2019} with the \T{switch} function being explicitly part of the \say{control structures} in the R~Language Definition~\cite{rcoreteam_language_2023a}. Because of the more prominent use, we focus on the \T{if} and \T{if}-\T{else} constructs first~(\ref{ssec:c-if}), before discussing the usage of other conditional functions~(\ref{ssec:c-fun}).

\subsubsection{If and If-Else}\label{ssec:c-if} Fundamentally, only around~\ThePercent[0]{4692}{4692+8359} of the \T{if}'s in research and \ThePercent[0]{938115}{938115+1952793} of the \T{if}'s in \cran sources contain an \T{else} branch, revealing no clear preference \ExactFisheR{
    research with    =              4692,
    research without =         4692+8359,
    cran with        =            938115,
    cran without     =    938115+1952793,
    with             = {if without else}%
}, with the majority of bodies only consisting of a single expression (see \cref{fig:if-types}). Both constructs are around twice as common in the \cran dataset when compared with the research scripts.

Furthermore,~\ThePercent[1]{65+3}{8349+4692} of~\T{if} and \T{if-else} constructs in the research and~\ThePercent[1]{2873+686}{1952793+938115} in the \cran dataset are unnecessary, in that their condition is hard coded to be either true or false \ExactFisheR{
    research with    =              65+3,
    research without =         8349+4692-(65+3),
    cran with        =            2873+686,
    cran without     =    1952793+938115-(2873+686),
    with             = {unnecessary if(-else)}%
}.

\subsubsection{Conditional functions}\label{ssec:c-fun}
Compared to the frequent use of~\T{if} and~\T{if}-\T{else} in the \cran dataset, the \T{ifelse} and \T{switch} functions are used much less frequently, on average~\TheFrac{116297+128}{\ProcessedCRANFiles} and~\TheFrac{29978+9}{\ProcessedCRANFiles} times per file respectively. For the research scripts, however, \T{ifelse} is used more, with on average \TheFrac{5347+0}{\ProcessedSocialFiles} occurrences per file, while \T{switch} is used only \TheFrac{45+0}{\ProcessedSocialFiles} times per file \SigEffMann{-18}{0.167942098059917}.

\begin{finding}
    \subfinding Most \T{if} and \T{if-else} constructs are used as simple conditionals with a single expression as the body.
    \subfinding Comparable control-flow functions like \T{ifelse} and \T{switch} are used much less frequently, with the exception of \T{ifelse} in the research dataset.
\end{finding}

\begin{figure}
    \centering
\begin{subfigure}{\linewidth}
    \centering\includegraphics[scale=1.3]{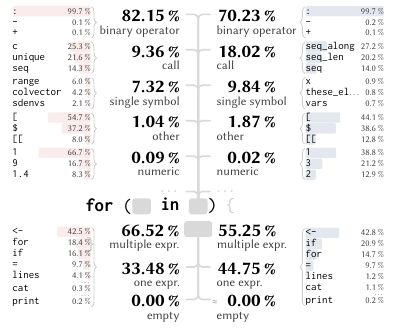}
    \caption{\TheFrac{13146}{\ProcessedSocialFiles} for loops per \ColorExample{@soc}~research file, \TheFrac{403091}{\ProcessedCRANFiles} per \ColorExample{@cran}~\cran file.\hfill\strut}
    \label{fig:loops-types-for}
\end{subfigure}\\[.65em]
\begin{subfigure}{\linewidth}
    \centering\includegraphics[page=2,scale=1.3]{plots/loops-types.pdf}
    \caption{\TheFrac{373}{\ProcessedSocialFiles} while loops per \ColorExample{@soc}~research file, \TheFrac{28500}{\ProcessedCRANFiles} per \ColorExample{@cran}~\cran file.\hfill\strut}
    \label{fig:loops-types-while}
\end{subfigure}\bigskip\\[.65em]
\begin{subfigure}{\linewidth}
    \centering\includegraphics[page=3,scale=1.3]{plots/loops-types.pdf}
    \caption{\TheFrac{49}{\ProcessedSocialFiles} repeat loops per \ColorExample{@soc}~research file, \TheFrac{3086}{\ProcessedCRANFiles} per \ColorExample{@cran}~\cran file.\hfill\strut}
\end{subfigure}

    \caption{Expression types used as the vector, condition, and body of R's explicit loop structures. With the research dataset on the left, and the \cran dataset on the right. The structure is similar to that of \cref{fig:assigned-types}.}
    \label{fig:loops-types}
\end{figure}

\subsection{Loops}\label{sec:result-loops}
The R~language offers three types of explicit loop structures~\cite{rcoreteam_language_2023}: \T{for}~(\ref{ssec:l-for-in}), \T{while}~(\ref{ssec:l-while}), and \T{repeat} loops~(\ref{ssec:l-repeat}). Furthermore, it offers \T{next} and \T{break} to manipulate the control flow in loops, which we address in \cref{ssec:l-break-next}, and other looping functions like \T{apply}, which we discuss in \cref{ssec:l-other}.
Overall we record a preference for explicit loops in research files when compared to the \cran sources \SigEffMann{-18}{0.292311027711108}, with a slight tendency for \T{for}-loops \gdef\ForLoopSig{{-18}{0.314139380886629}}\expandafter\SigEffMann\ForLoopSig.%

\subsubsection{For Loops}\label{ssec:l-for-in} R~offers only one kind of \T{for}-loop which iterates over all elements of a vector~\cite{rcoreteam_language_2023a}: \T{\textbf{for}(var \textbf{in} vector) body}. However, analyzing the vectors used as presented in \cref{fig:loops-types-for}, shows that most vectors~--- \ThePercent{10769+176+74+73}{13146} in the research and \ThePercent{282141+19738+14698+10174+494}{403091} in the \cran dataset \ExactFisheR{
    research with    =   10769+176+74+73,
    research without =             13146-(10769+176+74+73),
    cran with        =  282141+19738+14698+10174+494,
    cran without     =    403091-(282141+19738+14698+10174+494),
    with             = {for int-iteration}%
}~--- are constructed in place using the colon function~\T{a:b}, its function equivalent \T{seq(a,b)}, or with its faster variants \T{seq\_along} and \T{seq\_len} that iterate over the length of another vector. In other words, the majority of \T{for}-loops take the form of integer loops. 

Interestingly, files in the research dataset contain, on average, around three times as many \T{for}-loops as files in the \cran dataset. \ThePercent{808}{13146}~of the \T{for}-loops in the research and~\ThePercent{26587}{403091} in the \cran dataset directly nest another \T{for}-loop, despite this being discouraged \ReviewLabel{burns-reason}in favor of R's usually faster to execute and easier to read vectorization operations~\cite[pp.~17]{burns_inferno_2011}.

Similarly to the unnecessary \T{if}-\T{else} constructs, a small number of \T{for}-loops do not loop:~\ThePercent{17}{13146} in the research and~\ThePercent{106+36}{403091} in the \cran dataset either use a constant as the vector or have no or a single expression as the body, that ends the loop (e.g., \T{break}). This is used, for example, to reduce the runtime of a script by effectively hardcoding a specific configuration \ExactFisheR{
    research with    =   17,
    research without =   13146-17,
    cran with        =  106+36,
    cran without     =  403091-(106+36),
    with             = {not looping for}%
}.

\subsubsection{While Loops}\label{ssec:l-while} R's while loop functions similar to that of other languages: \T{\textbf{while}(condition) body}.
With~\qty{85.25}{\percent} in the research and~\qty{77.89}{\percent} in the \cran dataset, binary operators are the most prominent feature of the while-conditions, as shown in \cref{fig:loops-types-while}.
While most bodies contain multiple statements, we found a relatively large number of bodies with just a single expression: \qty{27.61}{\percent} in the research and \qty{13.04}{\percent} in the \cran dataset \ExactFisheR{
    research with    =   103,
    research without =   373-103,
    cran with        =  3715,
    cran without     =  28500-3715,
    with             = {single expr. while}%
}.
For both datasets, the majority of expressions in a single-expression-body are assignments (e.g., to retry random samples until a given predicate is met).

Only~\ThePercent{3}{373} of the while loops in the research and~\ThePercent{9+51}{28500} in the \cran dataset are trivially unnecessary, in the sense that they either have a constant \T{FALSE}~condition or the body consists only of a \T{return} or \T{stop} statement.

\subsubsection{Repeat Loops}\label{ssec:l-repeat}
Compared to the other loop constructs in~R, the \T{repeat} loop is used much less frequently. Furthermore, we encountered only one empty (and hence trivially endless) loop that appears in a \cran unit test.
Furthermore, while single-expression-bodies mostly contain assignments in the case of \T{for} and \T{while} loops, this does not hold for repeat loops.
Instead, the only expression used like that in repeats are try-expressions for the research dataset~--- this is reasonable given the fact that those loops would have no end condition otherwise.
The \cran dataset has a little bit more variety, but still, only~\qty{6.3}{\percent} expressions are assignments.

\subsubsection{Break and Next}\label{ssec:l-break-next}
\T{break} and \T{next} are used relatively rarely in our dataset, with only \num{199}~instances of \T{break} and \num{278}~of \T{next} in research code (i.e., \TheFrac{199+278}{13146+373+49} occurrences per loop). Similarly for the \cran dataset with only~\TheFrac{18833+57+532+10233+26+467}{403091+28500+3086} occurrences per loop.

\subsubsection{Looping Functions}\label{ssec:l-other}
Besides the explicit loops, R~offers several functions which hide looping,
with \T{lapply}~(\ThePercent[0]{466}{\ProcessedSocialFiles} of research, \ThePercent[0]{39666}{\ProcessedCRANFiles} of \cran files),
\T{apply}~(\ThePercent[0]{571}{\ProcessedSocialFiles}, \ThePercent[0]{30739}{\ProcessedCRANFiles}),
and \T{sapply}~(\ThePercent[0]{415}{\ProcessedSocialFiles}, \ThePercent[0]{31564}{\ProcessedCRANFiles}) being by far the most common. Overall they amount to around {%
    \edef\cranapplycount{\fpeval{137887+98960+100934+17383+9760+7613}}
    \edef\researchapplycount{\fpeval{2237+1945+3614+16+1372+99+1485}}%
    \TheFrac{\researchapplycount}{\ProcessedSocialFiles} and \TheFrac{\cranapplycount}{\ProcessedCRANFiles} occurrences per research and \cran file respectively.
}

\begin{finding}
    \subfinding Over~\qty{80}{\percent} of \T{for} loops operate on range vectors constructed in place. They are more than three times more frequent in research scripts.
    \subfinding Compared to the number of loops, \T{break} and \T{next} are used very rarely.
    \subfinding R's apply functions are used roughly as often as explicit~\T{for} loops.
\end{finding}

\subsection{Function Definitions}\label{sec:result-fun-def}
R~functions are defined using the \T{function} \ReviewLabel{fn-keyword-quotes}\enquote{keyword} (which can be, theoretically, redefined by the user), followed by the parameter list and the body. To give a function a name, it has to be assigned. For example, with one of the assignment operators presented in \cref{sec:result-assignments}. In the research dataset, we find \num{11616}~function definitions in~\NumAndPercent{1793}{\ProcessedSocialFiles} files. Only around half~(\ThePercent{6591}{11616}) are assigned to a name with the others used directly, for example, as parameters for higher-order functions.
The alternative function symbol~\say{\T{\textbackslash}} introduced in R~4.1 (released May~18, 2021) is not used in any of the research files. 
In contrast, we find function definitions in \NumAndPercent{236563+16729+1994}{\ProcessedCRANFiles} of all \cran files, \ThePercent{886244+35481+3681}{1236126+67202+5835} of which are assigned to a name~--- excluding those assigned using functions like \T{assign} or \T{setGeneric} (see \cref{ssec:a-used-fun}). Only~\ThePercent{814 + 170 + 6}{1236126+67202+5835} of definitions use the new function symbol~\say{\T{\textbackslash}}.

\subsubsection{Special Function Names}

Looking at special function names reveals that the hook functions~\T{.onAttach}, \T{.onLoad}, and \T{.on\-Unload} are the~\nth{5},~\nth{6}, and~\nth{15} most-used function names in the \cran dataset, defined in \ThePercent[1]{1807+4+27+5}{19450} of packages. These functions, are called implicitly when a package is attached, loaded, or unloaded, and can be used to, for example, initialize the state of the package~\cite{wickham2023r}.
They are completely unused in the research files.

\subsubsection{Non-Conventional Functions} \label{ssec:non-conv-fdef}If a function name starts and ends with the~\T{\%}-symbol,~R treats it as a so-called \say{special} infix operator, with the most prominent example probably being the pipe operator~\T{\%>\%} from the \textit{magrittr} package~\cite{magrittr}. While the \cran dataset defines more than~\num{800} different infix functions (for a total of \num{1964}~definitions), the research dataset only defines~\num{3} different names a total of \num{10} times: \T{\%nin\%}, \T{\%!in\%}, and \T{\%||\%}.
Besides these special infix-functions,~R allows to redefine all other infix-operators (like~\T{<-} and~\T{==}) as well. However, we found no instance of this in the research dataset and only a few in the \cran dataset, with the most prominent being the colon operator~\T{:}, redefined~\num{6} times in~\num{5} files.
Similarly, R's capability to redefine braces, brackets, and parenthesis, as well as control flow structures like~\T{if} or~\T{for}~\cite[Sec.~6.8]{wickham_advanced_2019} is unused in the research dataset and only used once in the \cran dataset.
Marginally more frequently used, are R's \textit{replacement functions}~\cite{wickham_advanced_2019} which allow calls on the target side of an assignment:
\begin{R}[literate={:bs:}{\textbackslash}1]
`f<-` <- :bs:(a, value){ a[1] <- value*2; a }
f(x)  <- 3
\end{R}
With a vector like \T{x <- c(1,2,3)}, this results in \T{c(6,2,3)}.
They all must have the special name \T{fun<-}, with \say{\T{fun}} being the name of the function. In the \cran dataset, we find~\num{1189} different replacement functions being defined a total of~\num{1346} times.

\begin{finding}
    \subfinding Minding the \ThePercent{814 + 170 + 6}{1236126+67202+5835} of definitions in the \cran dataset, R's alternative function symbol~\say{\T{\textbackslash}} is unused.
    \subfinding Around~\qty{9}{\percent} of \cran packages define hook functions (e.g.~\T{.onAttach}). Besides that, no function name is commonly defined or overwritten.
    \subfinding Around half of all function definitions are assigned to a name, slightly more so in the \cran dataset \ExactFisheR{
        research with    =   6591,
        research without =   11616-6591,
        cran with        =  886244+35481+3681,
        cran without     = 1236126+67202+5835-(886244+35481+3681),
        with             = {fun def assigned}%
    }.
    \subfinding Neither R's infix operators and its replacement functions, nor its ability to redefine braces, brackets, parenthesis, or control flow constructs are used in the research dataset.
\end{finding}

{\def\fn#1#2#3{\T{\strut\small#2}&\sisetup{round-pad=true,round-precision=1}\ThePercent[1]{#3}{#1}}
\def\fnS{\fn{\ProcessedSocialFiles}}
\def\fnRD{\fn{\ProcessedCRANDefaultFiles}}
\def\fnRE{\fn{\ProcessedCRANExampleFiles}}
\def\fnRT{\fn{\ProcessedCRANTestFiles}}
\setbox\CranBlob=\hbox{\ColorExample{@cran}}
\setbox\SocialBlob=\hbox{\ColorExample{@soc}}
\begin{table}
    \caption{Top five most called functions (used by \% of files).}\label{tab:fns}\vspace*{-.5\baselineskip}
\resizebox\linewidth!{%
\begin{tabular}{*4{l@{\hskip2pt}r}}
\toprule
    \multicolumn{2}{l}{\usebox\SocialBlob~research} & \multicolumn{2}{l}{\usebox\CranBlob~\cran~default} & \multicolumn{2}{l}{\usebox\CranBlob~\cran~{example}} & \multicolumn{2}{l}{\usebox\CranBlob~\cran~test} \\
    \cmidrule(r){1-2} \cmidrule(lr){3-4} \cmidrule(lr){5-6} \cmidrule(l){7-8}
    \fnS{c}{3547}          & \fnRD{c}{155871} &  \fnRE{library}  {2439} & \fnRT{test\_that}{54217} \\
    \fnS{library}{3099}    & \fnRD{length}{129106} &  \fnRE{c}{2401} &   \fnRT{c}{44497} \\
    \fnS{read.csv}{1966}   & \fnRD{list}{114712} &  \fnRE{list}{1386} &  \fnRT{expect\_equal}{36560} \\
    \fnS{length}{1830}     & \fnRD{stop}{90282} &  \fnRE{plot}{646} &    \fnRT{context}{27587} \\
    \fnS{list}{1731}       & \fnRD{is.null}{87798} & \fnRE{print}{602} & \fnRT{library}{40684} \\
\bottomrule
\end{tabular}}
\end{table}
}
\subsection{Function Calls}\label{sec:result-fun-call}
Internally, essentially everything in~R is a function call, including \T{for}-loops and parenthesis (cf.~\cref{ssec:non-conv-fdef}). Although it is certainly beneficial for static program analyzers to be aware of this and therefore treat everything as a call to a function, we use this section to discuss \enquote{conventional} calls~--- i.e., those in prefix or replacement form~\cite{wickham_advanced_2019}~--- and their consequences.
Specifically, we focus on R's reflective functions which, for example, allow the modification of functions at run-time~(\ref{ssec:c-reflective}) and calls using R's foreign function interface~(\ref{ssec:ffi}).
For an overview of the most-called functions, refer to \cref{tab:fns}. Interestingly, even though common testing functions\footnote{Like the \T{expect\_*} functions of the \T{testthat} package.} amount to \ThePercent[1]{1415062}{\AllCallsCRAN} of \textit{all} calls (\ThePercent[1]{1410247}{\AllCallsCRANTest}~in the test files) in the \cran dataset, none appear in any of the research files.

\subsubsection{Reflective Functions}\label{ssec:c-reflective}
R~offers a wide variety of reflective functions. \T{eval} evaluates code given as a string and has been previously shown to be in widespread use in R~packages often to provide functionality similar to macros in other languages~\cite{goel2021we}. We identify~\NumAndPercent{44}{\ProcessedSocialFiles} research and~\NumAndPercent{12203}{\ProcessedCRANFiles} \cran files which directly use~\T{eval} \ExactFisheR{
    research with    =  44,
    cran with        =  12203,
    with             = {direct eval}%
}.
R's~\T{body}, \T{formals}, and \T{environment} functions allow the access and manipulation of the body, parameters, and environment of a function.
While amounting to \ThePercent{2455+7270+27031}{\AllCallsCRAN} of all calls in the \cran dataset, they are only used by~\num{1},~\num{5}, and~\num{12} research files respectively.\footnote{Only four files change the formals of a function, either to assign default values or to clear R's~\say{\T{...}} parameter. Similarly, only four files manipulate a function's environment, each time to work with a copy.}
Other reflective functions are used infrequently as well:
{\def\SCR#1#2#3#4{~(\ThePercent[1]{#1}{\ProcessedSocialFiles}#3,~\ThePercent[1]{#2}{\ProcessedCRANFiles}#4)}%
\def\SCL#1#2{\SCR{#1}{#2}{~research}{~\cran files}}%
\def\SC#1#2{\SCR{#1}{#2}{}{}}%
\T{parse}\SCL{27}{6039} and \T{deparse}\SC{20}{9386}, which allow to convert strings into R~expressions and vice-versa, for example, to automatically generate labels in a plot.
\T{sub\-sti\-tute}\SC{34}{10701}, which does not evaluate an expression but allows to replace symbols in it and \T{quote}\SC{9}{3442}, which does not evaluate its argument.
} In general, we record a slightly increased usage of reflective functions in the \cran dataset \SigEffMann{-18}{0.0902114258940256}.

While only~\ThePercent[1]{2689+318}{\ProcessedCRANFiles} of \cran files make use of R's \T{load} and \T{attach} function, \ThePercent[1]{407+93}{\ProcessedSocialFiles} of the research files use it to load previously stored R~objects from a binary format \ExactFisheR{
    research with    =  407+93,
    cran with        =  2689+318,
    with             = {load/attach}%
}.

\subsubsection{Foreign Function Interface}\label{ssec:ffi}
R's foreign function interface allows to call compiled~C/\cpp and Fortran functions directly from R. While the \cran dataset contains~\NumAndPercent{2380}{\AllCallsCRAN} calls to Fortran, and~\NumAndPercent{43662+6351+333}{\AllCallsCRAN} calls to C/\cpp, the research files contain no call to Fortran and~\num{3} calls to C/\cpp. On manual investigation, all three appear to be copy-and-slightly-modified snippets of \cran code.

\begin{finding}
    \subfinding The research dataset does not make use of common testing libraries or functions, while test-functions amount to \ThePercent[1]{1410247}{\AllCallsCRANTest}~of all calls in \cran test files.
    \subfinding Even though R offers many reflective functions, the most used functions \T{eval} and \T{substitute} only appear in~\(\approx\){\qty3\percent} of \cran files (compared with \(\approx\)\qty1\percent\ of research files). However, \ThePercent[1]{407+93}{\ProcessedSocialFiles} of research files make use of R's \T{load} and \T{attach} functions to load previously stored R~objects.
    \subfinding The research dataset contained almost no explicit calls to C/\cpp and Fortran, with all three calls appearing to originate from \cran functions.
\end{finding}

\subsection{Packages}\label{sec:result-packages}

R~allows to load packages using either \T{library}, \T{require}, \T{load}-, \T{require}-, or \T{attachNamespace}. Furthermore, they can be directly accessed using~\bR{::} or~\bR{:::}, with the latter one allowing to access internal (i.e., not-exported) names as well~\cite{rcoreteam_language_2023a}.
In this section, we start by discussing the most used packages in \cref{ssec:p-used} before dealing with loading patterns and the alternative import methods of the \T{roxygen2} package~\cite{roxygen2} in \cref{ssec:p-loading}.

{\def\pkg#1#2#3{\T{\strut\small#2}&\sisetup{round-pad=true,round-precision=1}\ThePercent[1]{#3}{#1}}
\def\pkgS{\pkg{\ProcessedSocialFiles}}
\def\pkgRD{\pkg{\ProcessedCRANDefaultFiles}}
\def\pkgRE{\pkg{\ProcessedCRANExampleFiles}}
\def\pkgRT{\pkg{\ProcessedCRANTestFiles}}
\setbox\CranBlob=\hbox{\ColorExample{@cran}}
\setbox\SocialBlob=\hbox{\ColorExample{@soc}}
\begin{table}
    \caption{Top five most used packages (loaded by \% of files).}\label{tab:packages}\vspace*{-.5\baselineskip}
\resizebox\linewidth!{\begin{tabular}{*4{l@{\hskip2pt}r}}
\toprule
    \multicolumn{2}{l}{\usebox\SocialBlob~research} & \multicolumn{2}{l}{\usebox\CranBlob~\cran~default} & \multicolumn{2}{l}{\usebox\CranBlob~\cran~example} & \multicolumn{2}{l}{\usebox\CranBlob~\cran~test} \\
    \cmidrule(r){1-2} \cmidrule(lr){3-4} \cmidrule(lr){5-6} \cmidrule(l){7-8}
    \pkgS{ggplot2}{1142} & \pkgRD{stats}{15682} &  \pkgRE{shiny}  {674} & \pkgRT{testthat}{14417} \\
    \pkgS{dplyr}{978}    & \pkgRD{dplyr}{12145} &  \pkgRE{ggplot2}{315} & \pkgRT{dplyr}   {3119} \\
    \pkgS{tidyverse}{740}& \pkgRD{utils}{10367} &  \pkgRE{knitr}  {278} & \pkgRT{tibble}  {1827} \\
    \pkgS{lme4}{584}     & \pkgRD{knitr}{10216} &  \pkgRE{dplyr}  {236} & \pkgRT{withr}   {1311} \\
    \pkgS{car}{414}      & \pkgRD{ggplot2}{7084} & \pkgRE{DT}     {70} &  \pkgRT{stats}   {1026} \\
\bottomrule
\end{tabular}}
\end{table}
}

\subsubsection{Most Loaded Packages}\label{ssec:p-used}
\cref{tab:packages} presents the top five most loaded packages, and reveals that even though \T{ggplot2}, \T{dplyr}, and \T{tidyverse} are the most loaded packages in the research dataset, none of them are loaded in more than~\qty{30}{\percent} of the files. Similarly, no single package is loaded in more than~\qty{20}{\percent} of the \cran files.
This is due to several reasons:
\begin{enumerate}[nosep,wide=0pt]
    \item R~packages can automatically load and expose other packages. For example, loading \T{tidyverse} loads \T{dplyr} and \T{ggplot2} as well.
    \item Functions like \T{source} allow loading other R~files in the context of the current file. Especially together with \T{testthat}, packages do not have to re-import the library for every test file but can test the complete package with a function like \T{test\_check}.
    \item Packages may provide custom package-loading functions (e.g., \T{easypackages::libraries}) which are not detected by our analysis. However, due to the rare occurrences of such functions during our manual investigation, we consider the impact to be negligible. 
\end{enumerate}

\subsubsection{Loading Patterns}\label{ssec:p-loading}
\ThePercent{3095}{\ProcessedSocialFiles} of research files use the \T{library} function to load their packages, with the second-most used loading function being \T{require} used in only \ThePercent{488}{\ProcessedSocialFiles} of files. \cran packages on the other hand tend to prefer the use of \bR{::}, with only \ThePercent{16164+22302+2439}{\ProcessedCRANFiles} using \T{library} (most of these calls originate from test files).
Neither \T{load}-, \T{require}-, nor \T{attachNamespace} are used very often, only in \ThePercent{163+40+0}{\ProcessedCRANFiles}, \ThePercent{6406+1722+45}{\ProcessedCRANFiles}, and \ThePercent{91+13+0}{\ProcessedCRANFiles} of \cran files respectively. In the research dataset, only \T{requireNamespace} is used in \ThePercent{18}{\ProcessedSocialFiles} of files with the usual use-case being of the following shape (i.e., to install a package if it is not present):
\begin{R}
if(!requireNamespace('X', quietly=TRUE))
    install.packages('X')
\end{R}
\vspace*{-.66\baselineskip}

\paragraph{:: and :::}\quad \ThePercent{971}{\ProcessedSocialFiles}~of all research files use the~\bR{::} operator at least once to load a package, compared with~\ThePercent{90455+23588+1136}{\ProcessedCRANFiles} files in the \cran dataset~\ExactFisheR{
    research with    =               971,
    cran with        =  90455+23588+1136,
    with             =              {::}%
}.
\bR{:::}~is used considerably less frequently, appearing only in \ThePercent{39}{\ProcessedSocialFiles} of research and \ThePercent{74+4215+1342}{\ProcessedCRANFiles} of \cran files. \ThePercent{28814}{4296+28814+316}~of \bR{:::} usages stem from test files, for example, to test internal functions of the own package.

\paragraph{Loading Multiple Packages} A relatively common pattern to load multiple packages which we encountered during the manual investigation of the datasets is the usage of an application function like \T{lapply} to load multiple packages at once, appearing in \ThePercent{39}{\ProcessedSocialFiles} of research and \ThePercent{228+40+5}{\ProcessedCRANFiles} of \cran files:
\begin{R}
pkgs <- c("ggplot2", "dplyr")
lapply(pkgs, require, character.only=TRUE)
\end{R}
\vspace*{-.8\baselineskip}

\paragraph{roxygen2} R's prominent \T{roxygen2} package for \enquote{in-line documentation}~\cite{roxygen2} offers the capability to import packages locally using tags like \T{@import} or \T{@im\-port\-From} using the \T{DESCRIPTION} file~\cite{wickham2023r}.
Even though used in only~\ThePercent{5}{\ProcessedSocialFiles} of research files, \T{@im\-port\-From} alone appears in \ThePercent{98+37993+43}{\ProcessedCRANFiles} of \cran files \ExactFisheR{
    research with    =  5,
    cran with        =  98+37993+43,
    with             = {importFrom}%
}.

\begin{finding}
    \subfinding There is no single package dominating all research or \cran files, although there is a trend of \say{\textit{tidyverse} packages} for research, and the \textit{testthat} package for \cran test files.
    \subfinding While most research scripts~(\ThePercent[1]{3095}{\ProcessedSocialFiles}) use \T{library} to import packages \ExactFisheR{
        research with    =  3095,
        cran with        =  16164+22302+2439,
        with             = {library to load package}%
    }, \cran packages tend to use~\bR{::} instead. \ThePercent{28814}{4296+28814+316} of all usages of the \bR{:::} operator, which allows to access internal names, happen in test files.
    \subfinding The import functionality of the \T{roxygen2} package is used primarily in \cran packages.
\end{finding}

\section{Discussion}\label{sec:discussion}
We use this section to address our research questions, handling~\ref{rq:main}, the common and lesser used features in \cref{sec:common-and-lesser-used-features},~\ref{rq:diff}, the differences between research scripts and packages in \cref{sec:differences-between-scripts-and-packages}, and~\ref{rq:cons}, the insights for static analysis tools in \cref{sec:insights-for-static-analysis-tools}.

\subsection{Common and Lesser Used Features}\label{sec:common-and-lesser-used-features}
\paragraph{Common} There is a clear trend of~\T{<-} and~\T{=} for assignments as well as redefinitions within the same file, revealing the prominence of imperative constructs, even though R~supports functional concepts. Moreover, around~\qty{1}{\percent} of files use the super assignment operators, for example, to provide stateful functions.
We hence recommend static analyzers to offer support for these features.
Corresponding assignment functions are used less but still occur in around \qty{4.5}{\percent} of \cran files.
In general, about half of them are assigned to a name, with several packages using common hook functions like~\T{.onLoad} which are called implicitly when a package is loaded.
While assignment functions and hooks can be safely ignored by static analyzers focusing on errors in research code, we recommend their support if package code is to be analyzed as well.
Observing on average more than two usages per research file, we can see a high trend of \T{for} loops and \T{if} constructs as well,~1/3 of which use an \T{else}-branch. Similarly, we find a high number of indexing operations using~\T{\$}, i.e., accessing a data frame using a constant.
While complete support for tracking data frames and, for example, available columns, is difficult, the high number of constant-based indexing operations suggests great value in supporting at least the simpler subset of access by constants.
For package loading, there is a clear preference for \T{library} and~\T{::}, both of which we recommend to support in static analyzers that incorporate packages into their analysis.

From R's vast reflective arsenal, especially~\T{load} appears relatively frequently in the research dataset, followed by~\T{eval} and \T{substitute} in the \cran files. Albeit full support for functions like \T{eval} is difficult, we recommend to, at least, detect these functions during the analysis, allowing for pessimistic abstractions.

\paragraph{Uncommon} Even though~R allows for features like \begin{inlist}
   \item special infix operators
   \item the redefinition of existing operators
   \item the redefinition of language constructs like \T{if}
\end{inlist}, all of those features are rarely used, with~R's replacement functions and special infix operators, being the most common. Hence, handling them is of no great priority.
Similarly rare is the usage of the \T{switch} functions (on average only in one of~12~\cran and one of~100~research files) and \T{repeat} loops (on average only in \qty{1}{\percent} of files), as well as~R's ability to prevent bindings from being modified (with just 2~occurrences in the research and~137 in the \cran dataset).
Furthermore, there are merely \NumAndPercent{3}{\AllCallsSocial} direct calls to the FFI in the research dataset, despite their importance for performance~\cite[Chp.~25]{wickham_advanced_2019}, which suggests that those calls are hidden in the respective libraries or due to a bias in our dataset.
Therefore, a static analyzer which focuses on helping researchers, can ignore such calls.

\subsection{Research Scripts and Package Code}\label{sec:differences-between-scripts-and-packages}
Comparing the results of the research scripts and the package code, we find the former to be on average three times as long, with a higher usage of the discouraged~\T{=} assignment operator, especially in combination with~\T{<-}.
Moreover, research scripts tend to use more~\T{for} loops \expandafter\BasicSigEffMann\ForLoopSig\space but fewer~\T{if} constructs (on average~\qty{62.6}{\percent} less).
Absent from research scripts as well are common testing functions, direct calls to the FFI, and most of R's reflective functions, with the only exception of~\T{load} and~\T{attach} (used in~\qty{12.2}{\percent} of research files).
To load packages, research scripts primarily use \T{library} \LoadFisher{library to load package}, while package code tends to use~\T{::} as well as the capabilities of the \textit{roxygen2} package.
In general, research scripts tend to have a slightly more unnecessary \T{for}-loops \LoadFisher{not looping for} and~\T{if} conditions \LoadFisher{unnecessary if(-else)}, for example, to \enquote{hardcode} a configuration.

\subsection{Insights for Static Analysis Tools}\label{sec:insights-for-static-analysis-tools}
We separate our insights, next to those collected as part of \cref{sec:common-and-lesser-used-features}, into two categories: \begin{inlist}
   \item things to bear in mind when processing R~code
   \item potential improvements for the existing \T{lintr} package as the major static analysis package for~R
\end{inlist}.

\subsubsection{Processing R~Code}
Even though~\qty{3}{\percent} of research files failed to parse, all syntax errors can be compensated for with a simple recovering mechanism (e.g., skipping the respective line).
Similarly, for the failed \cran files, encoding errors can be compensated by explicitly reencoding the file, or updating the \T{xmlparsedata} package to make it more robust. All documentation errors can be handled by checking for the presence of documentation commands and, if desired, using a dedicated function like \T{tools::parse\_Rd}.

\ReviewLabel{reflection-semantics}\paragraph{Handling Reflection} After parsing the file, we consider handling R's reflective capabilities to be the hardest part (as it requires heavy-weight abstract interpretation, tracking scopes,~\ldots).
Our findings indicate, that it is sufficient to focus on \begin{inlist}
   \item the \T{load} and \T{attach} functions, which are used in \ThePercent[1]{407+93}{\ProcessedSocialFiles} of research files
   \item \T{parse}, \T{deparse}, \T{eval} and \T{substitute} appearing in~\qtyrange{1}{3}{\percent} of \cran and around~\qty{1}{\percent} of research files
   \item R's assignment functions like \T{setMethod} used in \ThePercent{4868+35+3214+7106+844}{\ProcessedCRANFiles} of \cran and \ThePercent{59+1}{\ProcessedSocialFiles} of research files
\end{inlist}.
However, handling those sufficiently is still difficult.
Although \T{substitute}'s implementation behavior of not evaluating expressions can be probably handled by treating the expression as a function call with all unbound variables acting as parameters (and~\T{eval} working as the corresponding call), the implementation semantics of \T{eval} are more difficult to handle and we are currently not aware of a solution which handles this correctly.
Moreover, the semantics of \T{setMethod} are fundamentally generic and require keeping track of the classes that are linked to the arguments when calling the function. Ultimately this concludes, that there is still a lot of research required to provide static analysis support for such features.\footnote{Although there is related work for Java, which may be applicable to~R as well~\cite{landman2017challenges,liu2017reflection}.}
Yet, it may be possible to cover a large set of real-world use cases by covering a small set of usage patterns, like the usage of \T{parse} and~\T{deparse} in the context of labels for graphical plots.

Even though identifying the name of a loaded package is relatively simple, identifying the precise version requested is not part of the load, although a package's \T{DESCRIPTION} file usually contains information on the minimum package versions required. However, we did not include \T{DESCRIPTION} files in our analysis, focusing only on R~code. To our knowledge, there is no existing tool that can automatically determine all viable package versions for a given R~script (e.g., based on the exported functions by the package), requiring analyzers to rely on the versions installed on the user's system (as the \T{lintr} package does) or to guess themselves.

\subsubsection{Improvements for \T{lintr}}
Although the current implementation of the \T{lintr} package, can already handle a variety of cases, we identified several scenarios that are not covered by current rules~\cite{lintr23}: \begin{inlist}
   \item even though its assignment linter can forbid super-assignments and enforce the use of~\T{<-} there is currently no mechanism that allows tackling the inconsistent use of assignment operators, which we found in~\qty{41}{\percent} of all research and~\qty{14}{\percent} of all \cran files
   \item when searching for unreachable code, \T{lintr} restricts itself to detecting \T{TRUE} and \T{FALSE} but ignores all other constantly false or true values (like \enquote{\T{0}})
   \item there is currently no \say{strict} mode provided which prevents the usage of complex, hard to understand, and potentially dangerous features like \T{eval} or functions which are openly discouraged like \T{assignInNamespace}, even though they appear in~\qty{1.1}{\percent} of research files
\end{inlist}.

\section{Threats to Validity}\label{sec:threats}
The following paragraphs discuss the reliability and reproducibility of our results~(\ref{ssec:reliability-and-reproducibility}), followed by a discussion of the internal~(\ref{ssec:internal-validity}) and external validity~(\ref{ssec:external-validity}) based on \citeauthor*{yu_threats_2010}~\cite{yu_threats_2010,mcdermott_internal_2011}.

\subsection{Reliability and Reproducibility}\label{ssec:reliability-and-reproducibility}

We provide a complete reproducibility package (see \cref{sec:data-availability}).
As no part of the automated analysis involves non-determinism, we consider our results to be reliable and reproducible. For our manual classification, we can only provide our classifications, usually reached in a group of two or three, for further criticism.

\subsection{Internal Validity}\label{ssec:internal-validity}
Our approach relies solely on the AST, including the dataflow extracted from it, without any additional information. We recognize the following \ref{internal-threats-last}~sources of error regarding our instrumentation:
\begin{enumerate}[nosep,wide=0pt,ref=\arabic*]
   \item Our XPath-expressions may be incomplete or contain errors, incapable of finding, for example, all assignments or symbols.
   We counteract this by a set of unit tests and extensive manual checks.  
   Moreover, we adapt our XPath expressions from those used by the R~languageserver~\cite{languageserver23} and the \T{lintr} package~\cite{lintr23}.
   \item Similarly, the analyzed dataflow can be wrong and therefore cause the extraction to, e.g., miss or over-count redefinitions. We compensate for this, by a vast set of unit tests and manual checks.
   \item We can not correctly handle redefinitions of assignment operators so that they no longer fulfill their original purpose. Consider \T{`<-` <- `*`}, which changes the meaning from \T{<-} to that of multiplication. Now, \T{2 <- 3} yields the value~\(6\).
   In theory, this affects all statistics.
   Yet, even after an excessive search, we found no problematic use of such a redefinition.
   \item Several R~functions (e.g., \T{parse} and \T{deparse}) allow to freely switch between a string and an expression. Hence we do not detect function calls that, e.g., happen only in combination with \href{https://www.rdocumentation.org/packages/base/versions/3.6.2/topics/eval}{eval} or definitions with \T{assign}. This affects all results, as all kinds of code can be hidden with a piece of code like the following: \T{eval(parse(text="x <- 2"))}.
   This, currently, can not be helped as trying to support all possible constructs leads down a never-ending rabbit hole of abstract interpretation (cf.~\cref{ssec:c-reflective}).

   \item \label{internal-threats-last}We can not always correctly identify (potentially) loaded packages. Consider the second example from \cref{ssec:p-loading}.
   We must detect that \T{require} is used as a package-loading function, repeatedly applied on a vector \T{pkgs} which in this example directly contains the package names.
   While this does not create false positives, it certainly creates false negatives. As part of our manual exploration of the dataset, we identified around \num{60}~occurrences that load packages comparable to the previous example (cf. \cref{ssec:p-loading}).
\end{enumerate}

\subsection{External Validity}\label{ssec:external-validity}

Without any previous study regarding variations in R~code, it is difficult to argue that our findings are representative of the usage characteristics of the R~programming language as a whole and it is for future studies to investigate this. However, with our dataset covering the sources of all \cran packages, we are confident that our results are representative of the usage characteristics of R~packages.

Nevertheless, our results are potentially biased by the chosen sources for the R~scripts as we only analyzed a relatively small set of research submissions originating from three sources and can not possibly know if they are representative of the whole \say{R~ecosystem} (e.g., scripts on GitHub). Additionally, we ignored other formats that allow R~code like RMarkdown, which allows for comments to be interspersed with the actual R~code (similar to Python notebooks).

\section{Conclusion}\label{sec:conclusion}

This paper presents our large-scale static analysis on more than 50 million lines of real-world R~code, with which we identified assignments with~\T{<-} and~\T{=}, \T{for} loops, \T{if} conditionals, and name-based indexing operations with~\T{\$} to be among the most commonly used features. We also find that research scripts tend to be longer than a package script, but seldom make direct use of the reflective capabilities of~R, except for loading previously stored R~objects.
We have shown, that even though~R~allows a lot of problematic constructs, only a few are used in practice. With this, we hope to encourage and support the development of more advanced static analyzers for R~programs.
Besides, we provide insights for static analysis tools and identify potential future work like: \begin{inlist}
   \item analyzing the impact of~R's implicit vectorization and its laziness on the precision of static analysis
   \item the potential of abstract interpretation, especially in combination with reflective functions
   \item using static analysis to prevent common semantic mistakes in R~code
\end{inlist}.

\section{Data Availability}\label{sec:data-availability}
We provide the replication package publicly~\cite{rep-package}. It contains the \begin{inlist}
   \item sources for our datasets
   \item a docker image to reproduce our experiments
   \item the source code of all tools and scripts
   \item all raw as well as all summarized results of our experiments
\end{inlist}.

\begin{acks}

We thank Jakob Pietron and the anonymous reviewers for
their comments and suggestions.
This work was partially supported by the \grantsponsor{dfg}{German Research Foundation (DFG)}{https://gepris.dfg.de/gepris/OCTOPUS}: \textit{\grantnum{dfg}{504226141}} and \textit{\grantnum{dfg}{453895475}}.
\end{acks}

\bibliographystyle{ACM-Reference-Format}
\bibliography{abbrev,references}

\end{document}